# Estimating the potential shift from conventional public transport to flexible services based on smartcard transactions

Nir Fulman[a], Maria Marinov[b], Itzhak Benenson[a]

fulmannir@gmail.com; maria@braude.ac.il; bennya@tauex.tau.ac.il

[a] Department of Geography and Human Environment, Porter School of Environmental Studies, Tel Aviv University, Tel Aviv, Israel
[b] Department of Industrial Engineering and Management, Ort Braude College, Karmiel, Israel

We assume that urban travelers may prefer flexible modes of transportation over conventional public transport (PT) for making non-routine trips, and estimate the potential for such modal switch based on a database of 63 million smartcard records of PT boardings made in Israel during June 2019. The behavioral patterns of PT users are revealed by clustering their boarding records based on the location of the boarding stops and time of day, applying an extended DBSCAN algorithm. Our major findings are that (1) conventional home-work-home commuters are a minority and constitute less than 15% of the riders; (2) at least 30% of the PT trips do not belong to any cluster and can be classified as occasional; (3) The vast majority of users make both recurrent and occasional trips. A linear regression model provides a good estimate ($R^2$ = 0.85) of the number of occasional boardings at a stop as a function of the total number of boardings, time of day, and land use composition around the location of trip origin. We conclude that the conventional PT may lose substantial urban ridership to the future flexible modes.

**Keywords**

Public transport; Smartcard; Trip clustering; Ride-hailing; Demand-Responsive Transport; Transportation Network Company

# 1. Introduction

Before the 2000s, motorized urban transport was largely stagnant. A traveler could choose between rigid public transport (PT) services such as buses, light rail, metro, and trains with fixed routes and time schedules, and private cars or costly taxies offering complete schedule and route flexibility. In the early 2010s, new demand-responsive transportation (DRT) modes started to appear, exploiting mobile apps to match users to vehicles (Cohen and Shaheen 2018). The new Ride-Hailing (RH) and vehicle-sharing services are flexible in terms of routes, stops, and schedules. Typically, these services are operated by Transportation Network Companies (TNCs) that coordinate vehicle fleets to balance the operation costs and prices/level of service. The new services are more expensive than PT and, typically, the prices of RH may be close to the price of regular taxis, yet users describe them as advantageous in terms of convenience, comfort, and safety (Rayle et al. 2016). As a result, since being introduced, the use of DRT services steadily grows (Graehler et al. 2019). The anticipated transition of DRT fleets to autonomous vehicles is expected to strengthen this tendency (Schaller 2021).

The introduction of DRT stimulated the hope that private car users would prefer this mode to their cars (Erhardt et al. 2021). This didn't happen. Instead, recent studies show that the areas served by ride-hailing and carpooling services experience a significant decline in PT ridership (Graehler et al. 2019; Erhardt 2022). Studies of the data accumulated by the TNCs demonstrate that their services are primarily used for leisure, errands, and other irregular and not work-related trips (Zhong et al., 2018; Tirachini & del Río 2019). Is there a relation between a traveler's switch from PT to DRT and the type of trip? The studies of PT ridership are mostly devoted to regular/repeating trips (Goulet-Langlois et al. 2018; Kieu et al. 2015; El Mahrsi 2014) and do not address this question. In this paper, we investigate the possible switch from regular PT to DRT analyzing a database of 63 million PT smartcard (SC) validations records in Israel from June 2019. The following Section 2 reviews SC-based studies regarding the occasional PT ridership and the demand for DRT services, Section 3 presents the dataset, and Section 4 is devoted to the PT ridership analysis. We discuss the results of this analysis and the policy implications of our findings in concluding Section 5.

# 2. Literature review

## 2.1. The (ir)regularity of public transport ridership

Typically, PT travel patterns are derived from datasets of onboard SC validations that list the IDs of the traveler, public transport line, boarding stop, and boarding time, and when available, the id of the alighting stop and time of alighting. The SC data are then matched to GTFS data (Google 2021) to identify the traveler's route and possible alighting stop in case it is not recorded. Numerous studies aim at classifying ridership patterns by clustering travelers based on the spatio-temporal characteristics of their trips, with a strong focus on the travel patterns of commuters or, more generally, frequent PT users. A close view of the riding patterns of the frequent users reveals that some of their trips do not follow their regular travel patterns.

Morency et al. (2007) showed that for 7,000 PT users in Gatineau, Canada, who boarded the bus, on average, once a day, 70% of the boardings were made at the three most frequently used stops. Yet, 70% of the stops that these travelers boarded were used not more than four times during the 10 months study period. El Mahrsi et al. (2014) studied a one-month dataset of bus rides in Rennes, France that contained 50K frequent users who used PT 10 or more days during that month. They classified these users into 16 groups based on the number of boardings they made in each hour of the day and demonstrated that 36% of them belonged to clusters that exhibit behavior not indicative of work commuting. Goulet-Langlois et al. (2016) analyzed bus and subway boarding data of 33K frequent users over 4 weeks in London, UK. Based on boarding location and time, they inferred individual activity sequences and classified the riders into 11 clusters. Only 45% of these sequences belonged to clusters with distinct home-work patterns, while the rest of the clusters exhibited substantial weekend ridership and trips to a variety of secondary locations. Analyzing the same data with entropy-based measures, Goulet-Langois et al. (2018) found that even for users with 40 or more monthly trips, some activity sequences included non-repeating trips to secondary locations, not typical of a working week.

The studies of smartcard records datasets show that many users ride PT infrequently and the few trips they make are unique. Ma et al. (2017) clustered PT travel patterns of 18M users in Beijing, China over one month. The cluster of commuters comprised only 10% of the riders who used PT 23 days a month on average and followed the same route 55 times. The largest cluster comprised 75% of the users who combined various activities – on average they used PT not more than four days during the month and used the same route not more than four times. Kieu et al. (2015) examined one million PT users who boarded the bus, train, and ferry over four months in Brisbane, Australia, and found that only 36% of them made more than 50 trips during this period. They clustered the users' trips by the proximity of the trips' origin and destination stops and by boarding time of day and showed that none of the trips made by 64% of the users belonged to a cluster.

A few papers compare daily patterns of regular and irregular trips. Kieu et al. (2015) and Ma et al. (2017) demonstrated that commuters exhibit distinct morning and evening ridership peaks in contrast to irregular users, who use PT uniformly throughout the day. The ridership patterns of the users belonging to young and senior riders are less regular compared to the patterns of the other users - students and seniors board more often than other riders at the stops that they used less frequently. In Morency et al. (2007), senior riders were the least regular in their spatio-temporal behavior, and in Kieu et al. (2015), the origins and destinations of the seniors' trips were the most varying among the groups of PT riders. Manley et al. (2018) examined the regularity of trips, rather than the regularity of users' travel patterns based on 640M bus and train rides in London over three months. Clustering boardings by the time of day and accounting for lines and stations, they found that only 35-40% of train boardings (made by as little as 17% of users) were regular. Ridership was two-four times more regular in the morning and afternoon peaks than at noon and in the evening. High regularity of train boarding origins was observed on the outskirts of London, while the regularity of the boarding origins in the city center was low.

To conclude, the studies of trip regularity for various datasets, by applying different clustering methods, consistently indicate that work-home commuting is only one option for PT network usage. An essential share of PT trips can be classified as "irregular", for which boarding time, stops, and routes are not repeated throughout the month. In this paper, we focus on these occasional trips considering them as candidates for switching to flexible modes.

## 2.2. The (ir)regularity of trips performed with flexible transportation modes

The TNCs' data and field surveys repeatedly show that ride-hailing is primarily used for occasional trips, with leisure being the main trip purpose. Rayle et al. (2016) studied an intercept survey from the spring of 2014 that assessed the use of RH in San Francisco. They demonstrated that in 67% of the cases, RH was used for social/leisure activities such as visiting friends and family, or attending a concert, 16% commuted to and from work, 4% were airport trips and 5% were other errands, like doctor appointments. Tirachini & del Río (2019) found that in Santiago de Chile, 55% of RH trips were for leisure, 17% for shopping and errands, 6% for health, and only 17% for work. Similar results are reported in other studies (Clewlow and Mishra, 2017, Henao, 2017, de Souza Silva et al., 2018). Studies of paratransit and dial-a-ride ridership (see review in Jain et al. 2017) also show that these modes are mainly used for non-work purposes, like shopping, social visits, leisure, and health activities.

## 2.3. Flexible transportation modes attract PT users

Initially, the TNC companies' services were seen as potential first and last-mile solutions that could supplement public transport (PT), thereby facilitating car independence (Erhardt et al. 2021). Over the past decade, however, evidence has accumulated that these services compete with PT services for riders, thus adding to VMT, emissions, and parking demand. Tirachini & del-Rio (2019), based on a survey of 2017 in Santiago de Chile, showed that 38% of the RH trips substitute PT trips, 39% taxis, and 16% cars, while only 4% of the RH trips combine it with other modes like PT/bicycle. Henao and Marshall (2018), based on surveys in Denver, USA, found that RH substituted 22% of the PT trips, 23% cars, 14% carpool, 12% bike/walk, 10% taxi, 12% stated they would not make the trip otherwise and only 6% of trips continued with an additional mode of transportation. In the Greater Boston region in 2017, the modal shift to RH was estimated as 41% from PT, 40% from car and taxi, and 12% from active transport (Gehrke et al., 2019). Schaller (2021) studied the change in ridership caused by TNCs in several US cities during 2014-2020 and found that the shift from PT is a common tendency. The portion of the TNC services users who, previously, used PT, ranged between 50% (Chicago, New York) to 59% (San Francisco, Boston), those who previously used taxis ranged between 23-35% and those switching from private cars ranged 15-18%. Several authors predict that when Shared Autonomous Vehicles (SAV) would become a reality, they will mostly attract PT trips (Schaller 2021; Lavieri and Bhat 2019; Krueger et al. 2016; Vosooghi et al. 2019).

Comfort, security, price, and travel time are considered the major reasons for switching from conventional public transport (PT) to flexible modes (Tirachini & del-Rio 2019) and the competition between RH and PT leads to a substantial drop in the PT ridership. Graehler et al.

(2019), based on data on PT ridership in several large American cities between 2002 and 2018, found that every year TNCs attract an additional 1-2% of the bus and heavy rail ridership. Erhardt et al. (2021) studied the databases of TNCs and PT use in San Francisco and discovered a 10% drop in PT ridership between 2010 and 2015. According to Agarwal et al. (2019), on days when RH drivers were on strike in February, 2017, the metro daily ridership in New Delhi increased by 2.4%.

To conclude, the studies of smartcard record databases, data from field surveys, and TNC companies indicate the following possible evolution of PT ridership: Some of the trips of traditional PT users are regular, while some are occasional and are not repeated during the month. The trips performed with DRT are mostly for occasional purposes, like leisure or doctor visits, and the use of the DRT services during the decade of its existence steadily grows at the expense of traditional PT. This paper focuses on the first part of the above statement, and in what follows, we identify and characterize the occasional PT trips in a national smartcard dataset. We consider these trips as the major source of spillover to flexible modes of transportation. We consider our methodology suitable for any geographical area where PT services are ample and the information about the travels is available at the individual level.

## 3. Data

The investigated dataset consists of the 63M records of smartcard (SC) ride validations in June 2019 over the entire Israeli public transport network. Smartcards were used in PT for paying for 90% of all trips that month, while the rest of the payments were made in cash. The Israeli SC system for buses is tap-on only and a ride is recorded when the traveler boards the bus. For a train ride, alighting is also recorded. The validated ride can include transfers and is limited to 90 minutes. On transfer, the SC should be validated again but is not charged. The system of prices and discounts is cumbersome but, overall, the cost of a PT trip in Israel is low and a non-discounted trip within the city is 6 NIS and between cities is 15-35 NIS, while the average monthly personal income in Israel is 12,000 NIS (CBS 2022).

### 3.1. SC validation information

When the SC is validated, the information recorded on the operator's database is as follows: User's unique ID (recoded for this study to protect privacy), payment agreement (Basic Fare Pass (BP), Prepaid Pass (PP)), profile (General, Elderly, Student, etc.) that determines traveler's overall discount, boarding stop ID, line ID, exact time of the onboard validation and whether the ride is a transfer within 90 minutes of the initial validation. In the case of a train, the trip record contains the ID of the boarding station, the exact time of entry to the station, the ID of the final station of a trip, and the exact time of exit from the final station. Because the SC is registered at the entrance to/exit from the station, the transfers between the train lines are not registered. To locate lines and stops in space we exploit the open GTFS dataset of the Israeli Ministry of Transport (Google 2021). According to this database, c.a. 3,000 bus lines and 19,000 stops were operating in Israel in June 2019. Accounting for 90% of the boardings, buses are Israel's main form of PT. The remaining trips were made by train, with the Israeli national railway system that connects 70 stations, and the Jerusalem Light Rail line.

The payment agreement can be of two types: (1) The Basic Fare Pass ("BP" below) is paid every boarding from an electronic purse, at the time of boarding, and is 20% cheaper than cash payment to the driver on boarding and is further discounted 50% for Pupils and Senior Citizens; (2) Prepaid Pass ("PP" below) is paid in advance and allows an unlimited number of PT rides over predetermined areas for a certain period. The most common is a monthly PP within a predefined region, typically within a certain metropolitan area. The price of a monthly prepaid SC card is close to forty single trips within the same area. Available Prepaid cards depend on the user's profile: The second most common PP card is a semester card available only to students. Other forms, such as weekly or annual PP cards, are much less common and, together with special free passes for soldiers and very minor categories, comprise about 5% of the rides.

The standard workweek in Israel is from Sunday to Thursday and most PT lines operate between morning and midnight during this period. The PT system is almost fully suspended between Friday afternoon and Saturday evening. That is why, in what follows we filter out the 7M weekend boardings and analyze 56M transactions collected over 20 working days of June 2019: 02-06/06, 10-13/06, 16-20/06, 23-27/06, and 30/06.

### 3.2. Data cleaning

The smartcard (SC) validations database demanded substantial cleanup in several respects:

*Mistaken or illogical records*: In some cases, two sequential boardings were made by a user in two or fewer minutes. We interpret these cases as (1) accidental boarding on the wrong line and switching to the proper one and (2) a single card validated for more than one person. In both cases, only the last boarding was preserved. The fraction of deleted boardings is 1.7%.

*Too many boarding per day:* Users who boarded 12 or more times on one or more of 20 working days were excluded from the analysis examined. We believe these passengers use public transportation as part of their job, like delivery or sales. The fraction of deleted travelers is 0.6%.

*Unrecorded transfers:* We have evaluated the database for unrecorded transfers between lines. Namely, for two consecutive boardings of the same day, the first at a stop *a* of a line $l_a$ at a time $t_a$ and the second at a stop *b* of a line $l_b$ at a time $t_b$, we identify stops that are directly reachable from *a* with the line $l_a$ and then, the stop *c* on $l_a$ that is closest to *b*. The ride time $t_{a \to c, ride}$ with the line $l_a$ from *a* to *c* is estimated based on the GTFS dataset and the walk time $t_{c \to b, walk}$ from *c* to *b* is estimated based on the aerial distance between *c* and *b*, assuming a high walking speed of 5 km/h. The maximum waiting time at a stop *b* is assumed to be 20 minutes. Consecutive boardings at *a* and at *b*, for which $t_a + t_{a \to c, ride} + t_{c \to b, walk} + 20 > t_b$ are then considered as a transfer trip, and boarding at *b* was excluded from the analyzed dataset. Applying this rule, the fraction of deleted boardings is 14%.

### 3.3. Selection of data for analysis

The number of records in the database, after performing all cleaning procedures was 47M, and they are further analyzed regarding the travelers' profiles and payment agreements. Five agreement types comprise 99% of all agreement types:

1. Periodic Pass (PP): Monthly, yearly, and student semester tickets - 12.2M records of 308K users.
2. Basic Fare Pass (BP): Prepaid multiple-entry, stored-value, and daily pass - 23.6M records of 2.2M users.
3. Mixed Fare: Users switching between agreements during June 2019 - 7.2M records of 167K users).
4. Weekly Pass: 0.1M records of 2K users.
5. Special Free Pass: Users riding for free, like soldiers and public transportation employees. 3.9M records of 226K users.

For the analysis below, we have selected the users of the two major agreement types, PP and BP: 35.8M records by 2.5M users.

The database describes seven user profiles:

1. General Users: The default smart card profile. 19.5M records of 1.39M users.
2. Elderly/Seniors: 6.2M records of 396K users.
3. Youth/Teenagers: 6.8M records of 522K users.
4. Students: 2.0M records of 94K users.
5. Mixed: Users switching between profiles during June 2019, 1.0M records of 54K users.
6. Disabled: 0.3M records of 23K users.
7. Others: Minor profiles like police officers or parliament members, 0.1M records of 8K users).

We have chosen for our analysis the users of the four major profiles: Regular, Elderly, Teenagers, and Students. The investigated dataset contains 34.7M records that describe the use of PT by 2.4M users during July 2019.

## 4. Analysis

### 4.1. General view of PT ridership in Israel

On a typical working day in June 2019, 800K users boarded public transport (PT) and made 1.7M trips. 100K (12.5%) of them used Periodic Pass (PP) cards and 700K (87.5%) used Basic Fare Pass (BP) cards. On average, Israeli travelers board 1.7M/0.8M = 2.1 times a day. The distribution of users by the number of boardings per day for the PP and BP agreement types is presented in Table 1, with 33% of users boarding once a day.

Table 1: PT Use by the number of boarding per day, for the Prepaid and Basic Pass users

| Rides/day | Total (800K, 100%) | PP holders (100K, 12.5%) | BP holders (700K, 87.5%) |
|:---:|:---:|:---:|:---:|
| 1 | 33% | 15% | 40% |
| 2 | 41% | 46% | 39% |
| 3 – 4 | 22% | 32% | 18% |
| 5 – 6 | 4% | 6% | 3% |
| 7 – 12 | 1% | 1% | <1% |
| Avg. | 2.1 | 2.52 | 1.93 |

The intensity of monthly PT ridership is different for the PP and BP holders (Figure 1). The majority of PP users board PT most workdays of the month, the average number of days of use is 15.6, and the share of travelers who use PT grows with the increase in the number of travel days and reaches 25% for those who use PT every working day. For BP holders, the average number of travel days is 5.5, the share of those who use PT every working day – 25%, is the highest, and this share monotonously decreases with the increase in the number of travel days.

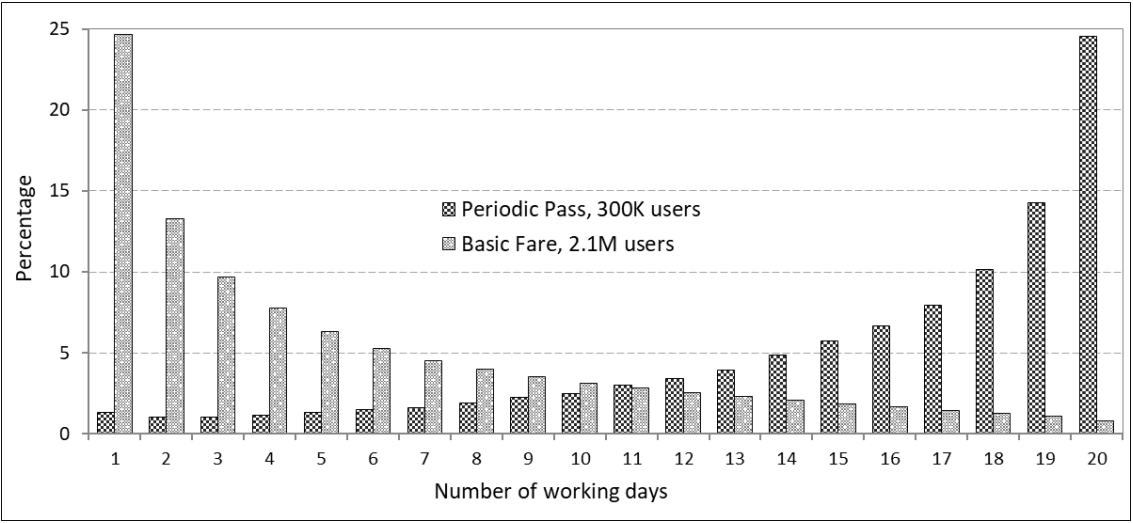

Figure 1: The distribution of the number of PT ridership days for PP and BP holders over the 20 working days of June 2019.

A similar tendency is characteristic of the number of monthly PT trips. The average number of boardings for PP card holders is 39.6, while for BP card holders it is 10.6. More than 60% of BP card owners board PT less than 10 times a month, and the median number of boardings for them is 7, while the median number of rides of PP owners is 38 (Figure 2).

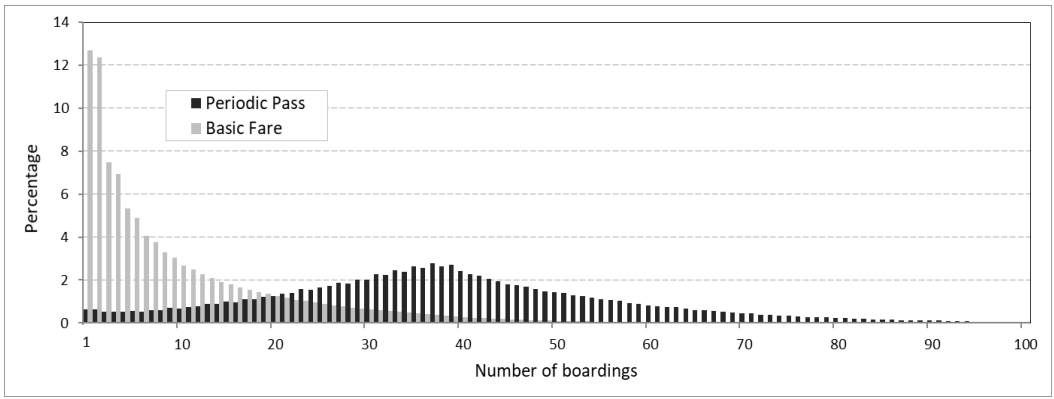

Figure 2: The number of PT rides on the working days in June 2019, for the PP and BP holders.

The same tendency is characteristic of the number of days of PT use per five working days of the week (Figure 3). The average number of working days when PP cardholders used the PT is 4.2, while for BP holders it is 2.2 days only.

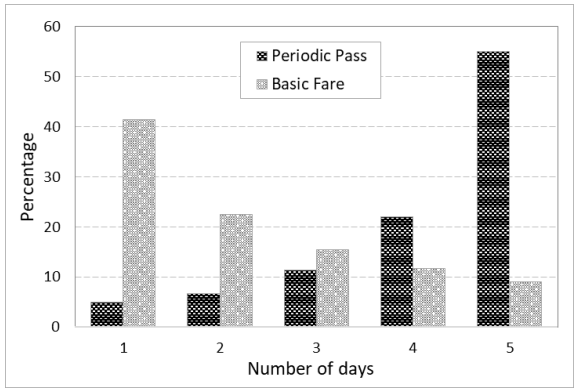

Figure 3: Weekly statistics of public transportation use for the PP and BP holders.

Table 2 specifies the statistics presented in Figures 1-3 by agreement type and user profile. The statistics of the major General profile are close to the overall average. Teenagers and Student BP holders ride more often than other BP holders, yet PP Teenagers and Students ride less than average for their agreement type.

| Table 2: Ridership statistics by four user profiles and two agreement types | | | | | | |
|---|---|---|---|---|---|---|
| Profile | Agreement | Users per day | Boardings per day | Days of use per 20 workdays of the month | Boardings per 20 workdays of the month | Days of use per working week |
| Elder | Prepaid | 22K | 2.7 | 16 | 41 | 4.1 |
| Elder | Basic | 108K | 2.0 | 5 | 10 | 2.0 |
| General | Prepaid | 65K | 2.5 | 16 | 40 | 4.3 |
| General | Basic | 400K | 1.9 | 5 | 10 | 2.2 |
| Student | Prepaid | 10K | 2.6 | 13 | 35 | 3.8 |
| Student | Basic | 22K | 2.0 | 7 | 14 | 2.4 |
| Youth | Prepaid | 3K | 2.3 | 15 | 35 | 4.1 |
| Youth | Basic | 170K | 1.9 | 7 | 13 | 2.6 |
| Total | Prepaid | 700K | 1.9 | 6 | 11 | 2.2 |
| Total | Basic | 100K | 2.5 | 16 | 40 | 4.2 |
| Grand total | ------- | 800K | 2.1 | 6.8 | 14.3 | 2.7 |

To sum up, PT ridership in Israel varies between travelers. 12.5% of them consider it worthwhile to purchase a Periodic Pass whereas the remaining 87.5% do not consider it worthwhile and hold a Basic Fare Pass. As we see from Table 3, this is the right decision: The PP holders' share of trips is 34% - almost three times higher than their share in the PT user population. BP owners perform the remaining 66% of the trips. The most striking finding is the high share of BP holders who use PT very infrequently: in July 2019, over 40% of them rode PT only one working day a week (Figure 3), up to three workdays a month (Figure 1) and no more than five times overall (Figure 2). Conversely, PP holders ride, on average, 4.2 workdays a week (Figure 3), 40 times per month (Figure 2), and 16 of 20 days in July 2019 (Figure 1). Yet 11% of them use PT up to two days a week (Figure 3), eight workdays a month (Figure 1), and 15 times a month (Figure 2).

### 4.2. Clustering Public transport trips

To understand the spatio-temporal regularity of public transport travelers' behavior we cluster their boardings by location of boarding stop and time of boarding. For this purpose, we applied a time-extended DBSCAN algorithm that extends the idea of the standard DBSCAN – the neighborhood of any point in the parameters' space should contain at least minPnt of other points (Ester et al. 1996) – by introducing additional temporal closeness between the boardings.

Formally, we apply spatial ($\epsilon_s$) and temporal ($\epsilon_t$) closeness thresholds: two rides boarded at stops, the distance between which is below $\epsilon_s$, are close spatially, and if the time interval between the rides is less than $\epsilon_t$ time units, they are close in the temporal sense (Figure 4). Applying DBSCAN, we thus require that, for it to be clustered, a boarding that is characterized by the location of the stop and the time of boarding should have at least minPnt other boardings within the ($\epsilon_s$, $\epsilon_t$) – neighborhood. Compared to the DBSCAN which is based on the location of the boarding stop only, accounting for the temporal closeness results in the splitting of some clusters constructed based on the boarding stop location into two or more clusters on basis of the time of boarding.

In what follows, we apply the extended DBSCAN with minPnt = 2, $\epsilon_s$ = 400m, and $\epsilon_t$ = 60 minutes. That is, the two trips are similar if they start at stops less than 400m from each other and the time interval between them is less than 60 minutes. The value of $\epsilon_s$ = 400m reflects the maximal walking distance riders are willing to walk to the nearest PT station (Canepa 2007). The value of $\epsilon_t$ = 60 is set based on the clusters of work commuting identified by El Mahrsi et al. (2014, see section 2.1), who found that over a month, a traveler's work trips are mostly boarded at the same hour of the day, every morning and evening. We intentionally employ the minimally possible value of minPnt = 2, to impose the minimally possible demand on a trip to be considered as regular: In our study, two similar trips per month are already considered as a cluster of regular trips. That is, the trip is considered as occasional if no other trip that month is spatiotemporally close to it. To examine the robustness of the results to the choice of three DBSCAN parameters - minPnt, $\epsilon_s$, and $\epsilon_t$, we repeat the analysis with different parameter values in the Appendix to this paper. Note that non-clustered trips may nonetheless represent a trip to a regular activity, like work or commerce, but we do not have data for investigating this possibility.

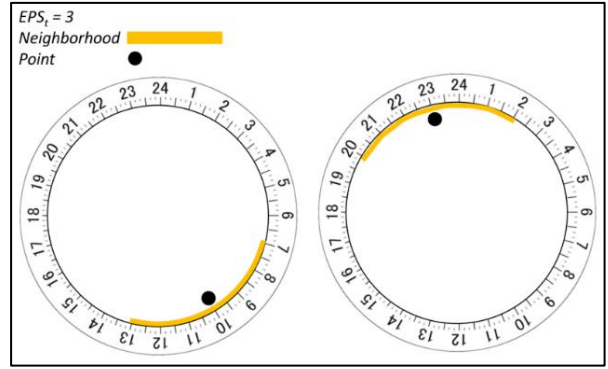

Figure 4: Temporal neighborhoods around boardings made at 10:00 and 23:00 for $\epsilon_t$ = 3 hours.

Typical spatio-temporal clusters of the June 2019 trips for a user with many - 58 boardings, and a user with a few – 10 boardings, are presented in Figure 5.

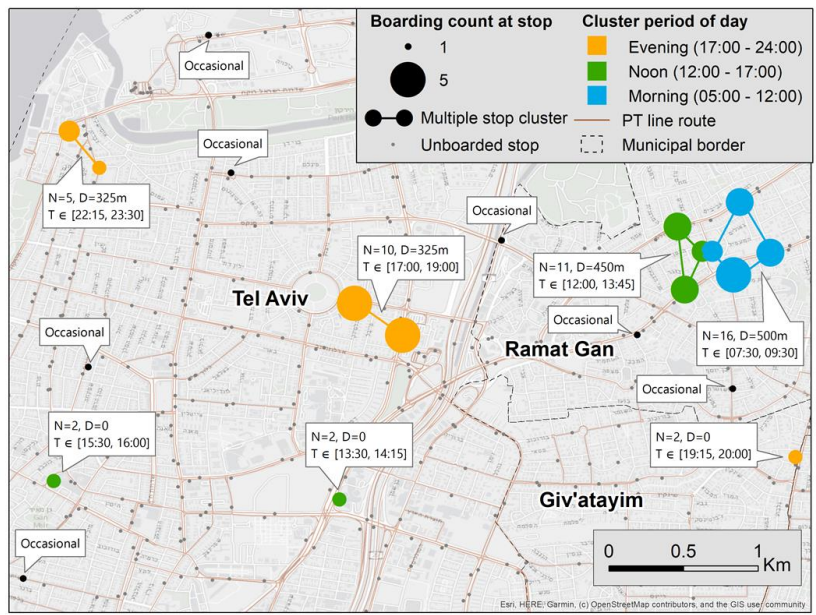

a

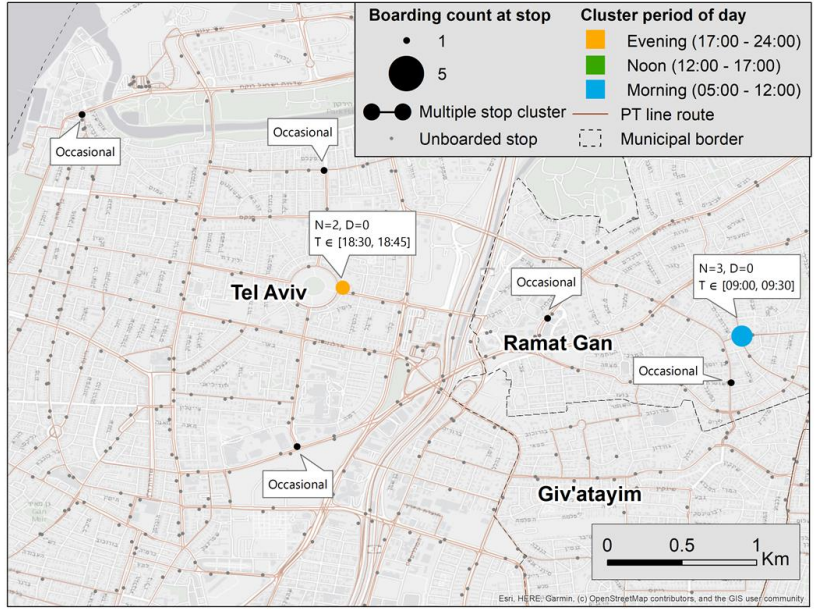

b

Figure 5. The clusters of boardings for two travelers of the "General" profile (a) a PP holder with 58 boardings in 18 working days, 48 clustered and 10 occasional, 4 outside the map scope, (b) a BP holder - 10 boardings in 5 working days, 5 of them clustered.

To conclude, interpreting results of the time extended DBSCAN, we classify a trip as *regular* if it belongs to one of the clusters, i.e., if at least one other trip started at a close PT stop at a close hour of the day. Otherwise, the trip is considered *irregular* or *occasional*.

### 4.3. Occasional trips by the users' groups and hours of the day

Table 3 presents the basic statistics of occasional trips, by agreement type and users' profiles.

Table 3: Boarding statistics by user profile and agreement type

| Profile | Agreement | Users | Boardings | Occasional, % |
|---|---|---|---|---|
| General | Prepaid | 199K | 7.9M | 23% |
| Elder | Prepaid | 70K | 2.9M | 26% |
| Student | Prepaid | 32K | 1.1M | 37% |
| Youth | Prepaid | 7K | 229K | 19% |
| **Total** | **Prepaid** | **307K** | **12.2M** | **24%** |
| General | Basic | 1.2M | 11.6M | 52% |
| Elder | Basic | 326K | 3.3M | 52% |
| Student | Basic | 62K | 832K | 53% |
| Youth | Basic | 516K | 6.5M | 49% |
| **Total** | **Basic** | **2.1M** | **22.5M** | **51%** |
| **Grand total** | --------- | **2.4M** | **34.7M** | **42%** |

According to Table 3, 42% of all trips are occasional with 51% among the BP holders and 24% among PP holders. Within the BP holders, Students are the most sporadic with 53% occasional trips, followed by General and Elder users with 52% and Teenagers with 49%. Within the PP holders, Students have 37% occasional trips, followed by Elders with 26%, General users with 23%, and Teenagers with 19%. The high percentage of occasional trips among Students possibly reflects their lifestyle and for Elders, it can be attributed to their 50% discounted and thus cheap card that makes it possible to ride unlimitedly.

Figure 6 shows that the share of occasional trips decreases drastically with the increase in the number of monthly trips made by the user. It declines from 60% for users with fewer than 10 monthly rides to 20% for users with 40 or more monthly boardings. BP users have only slightly higher shares of occasional trips than PP users with similar numbers of monthly trips.

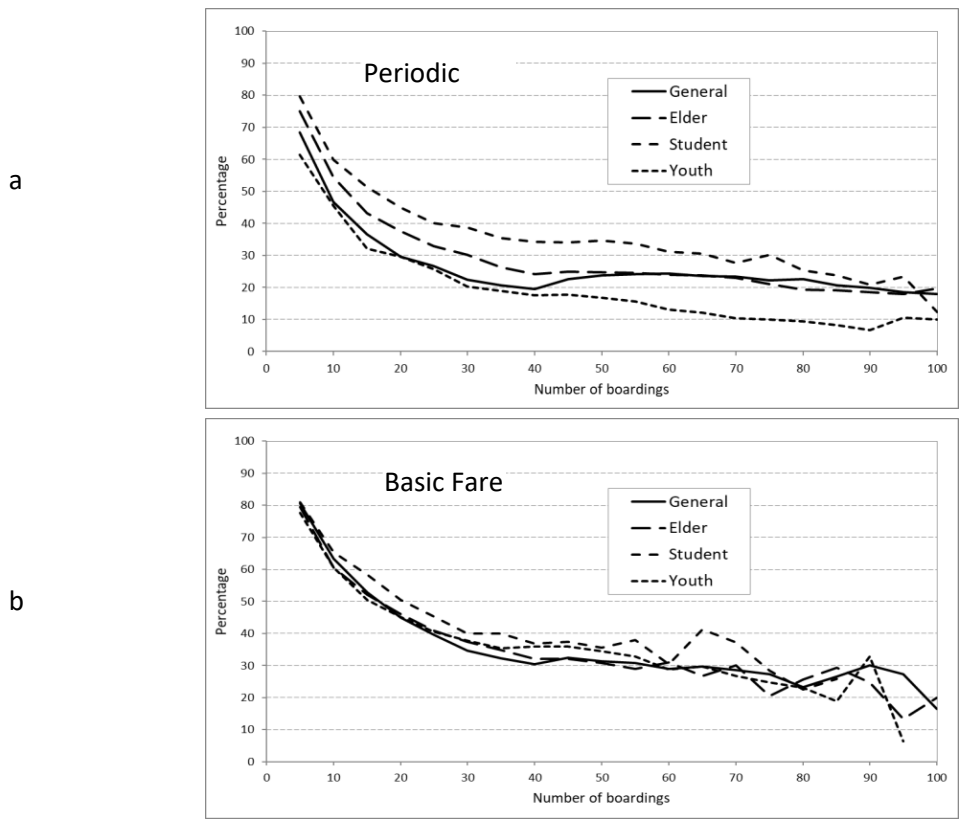

Figure 6: The share of occasional boardings by the users' agreement type, as dependent on the number of rides in working days of June 2019, for the holders of BP (a) and PP (b).

The average number of clusters increases linearly with the number of monthly boardings for both agreement types (Figure 7a) and the different user profiles (Figure 7b). The distribution of cluster sizes is far from uniform (Figure 7c-d): to remind, most BP users make less than 15 monthly boardings (Figure 2) and most of their trips are irregular (Figure 6). Typically, one large cluster includes half or more of their regular boardings (Figure 7c). PP holders make typically 20-60 monthly boardings (Figure 2), most of which are clustered (Figure 6). Most of these users have 1-4 large clusters possibly representing home and work activities, and smaller ones for more occasional trips (Figure 7d).

a

b

c

d

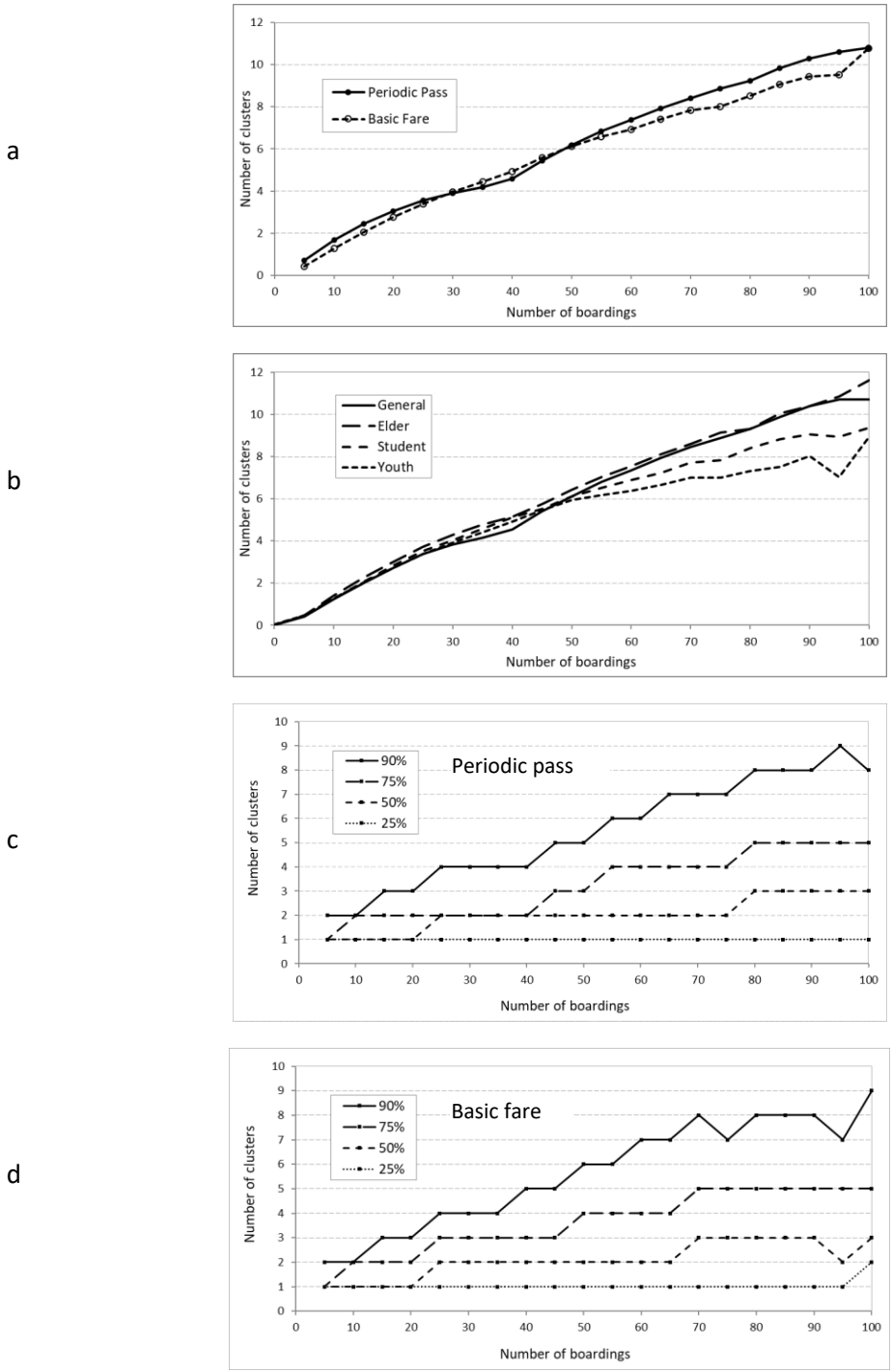

Figure 7: The average number of clusters by agreement type (a) and user profile (b) and the number of clusters that comprise a given percentage of boardings for the PP (c) and BP holders (d) as dependent on the total number of boardings.

The parameters of the extended DBSCAN algorithm $\epsilon_s$ = 400m and $\epsilon_t$ = 60 minutes define the spatio-temporal size of the clusters. Table 4 presents the clusters' average spatial and temporal dimensions: the time difference between the earliest and latest boarding in a cluster and the maximal diameter of the convex hull of the clusters' stops, as dependent on the number of boardings in a cluster.

Table 4: Clusters' average and standard deviation of the spatial and temporal dimension as dependent on the number of boardings in a cluster

|  | Time interval (minutes) | | Spatial diameter (meters) | |
| --- | --- | --- | --- | --- |
| Cluster size | Prepaid Pass | Basic Pass | Prepaid Pass | Basic Pass |
| 2 | 24 (STD = 18) | 24 (STD = 18) | 80 (STD = 128) | 83 (STD = 129) |
| 4 | 48 (STD = 36) | 48 (STD = 36) | 151 (STD = 192) | 158 (STD = 191) |
| 8 | 72 (STD = 48) | 78 (STD = 54) | 217 (STD = 251) | 229 (STD = 243) |
| 16 | 78 (STD = 66) | 84 (STD = 66) | 225 (STD = 283) | 244 (STD = 284) |

Overall, the clusters' dimension increases with the increase in the number of boardings in a cluster, up to eight boardings, and then stabilizes. The average cluster spatial diameter is only 135m because many clusters consist of multiple boardings made at the same stops and their spatial diameter is zero. The clusters of PP holders are slightly more compact than BP users, as can be expected (Figure 7).

The share of occasional trips varies widely by the time of day (Figure 8). It has a clear peak between 10:00 - 12:00 and declines later in the afternoon, with a typically higher percentage for BP cardholders compared to PP users. Teenagers' pattern is different from other groups' patterns, probably reflecting daily school schedules.

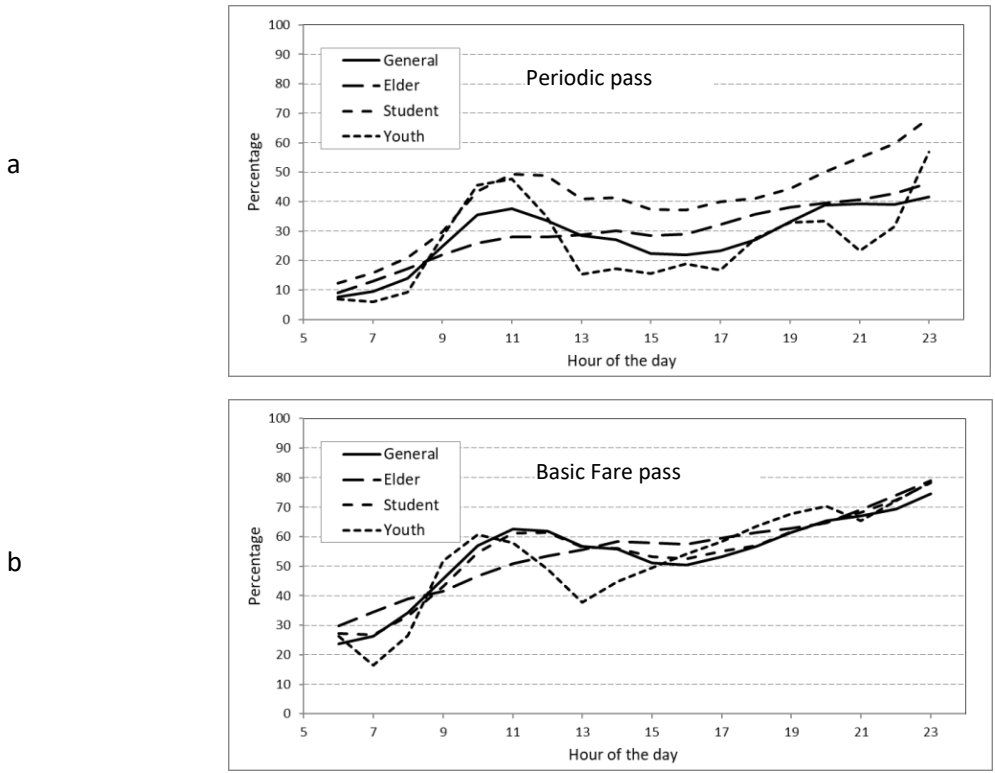

Figure 8: The share of occasional rides for the users of different profiles by the hour of the day: PP holders (a), BP holders (b).

To sum up, the spatio-temporal clustering of boardings reveals that occasional trips comprise 42% of all trips (Table 3). Note that the share of occasional trips for the BP holders is higher than for the PP holders (Table 3). The share of occasional trips drops with the increase in the number of monthly boardings, from 60% for users with fewer than 10 monthly rides to 20% for users with 40 or more monthly boardings. Ridership follows the daily pattern of activity, and the morning and afternoon trips representing the trip to and back from work are more regular than the trips in the afternoon and after work time (Figure 8).

To recognize the relationship between the share of occasional trips and PT users' activities let us investigate the variability of this share by bus lines and stops.

### 4.4. Occasional trips by lines and stops

The share of occasional trips varies greatly by lines (mean = 41%, STD = 24%) and stops (mean = 46%, STD = 24%), indicating that different lines and stops serve different purposes. Expectedly, the share of occasional trips for the Basic Fare Pass holders is higher and varies more than that for the Periodic Pass holders (Figure 9). The shares of the BP holders are distributed symmetrically around the mean, for all user profiles (Figure 9b, d). For the PP holders of the General and Elder profiles the distributions are also symmetric, but with essentially lower averages than for the BP holders (Figure 9a, c). The same distributions for the Students and Youth PP cardholders are different: Students have a higher share of irregular ridership compared to other groups, while the shares for Youth are relatively low.

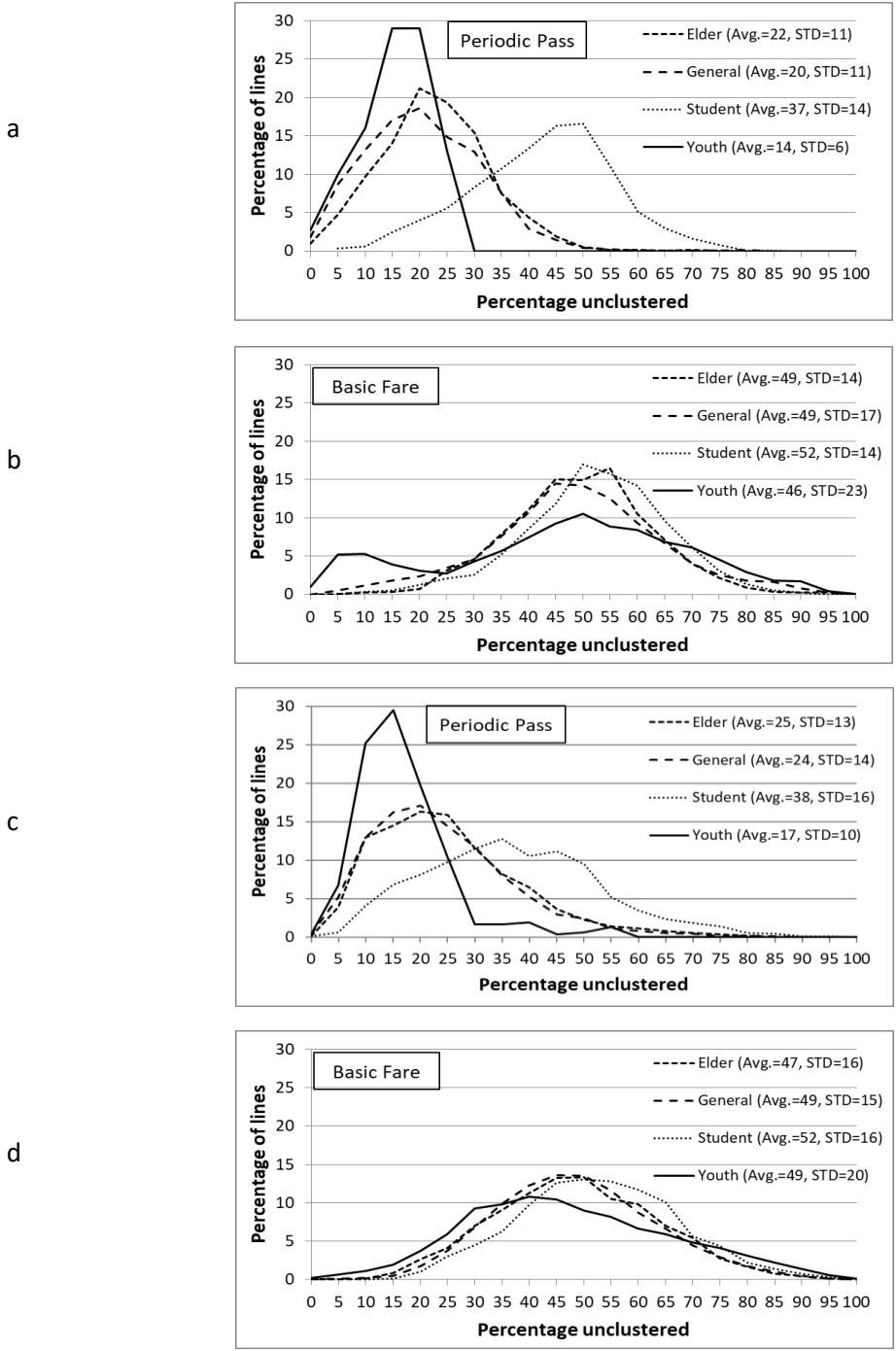

Figure 9. The distribution of the share of occasional trips by lines (a, b) and by stops (c, d).

To sum up, the share of occasional trips varies substantially by bus lines and stop, suggesting they are used for different types of activities. Some lines and stops have a remarkably high share of occasional trips.

## 4.5. Spatial variation of the share of occasional trips
### 4.5.1. Correlation over stops of the same line

Let us estimate the correlation of the daily shares of occasional boardings at the stops of the same line. The shares of occasional boardings at two sequential stops are strongly and positively correlated with r ~ 0.7 ($p < 0.01$) and with the increase in distance between stops the correlation decreases yet remains high at $r = 0.4$ (Figure 10) even for large lags. This trend is repeated for each user profile and period of the day.

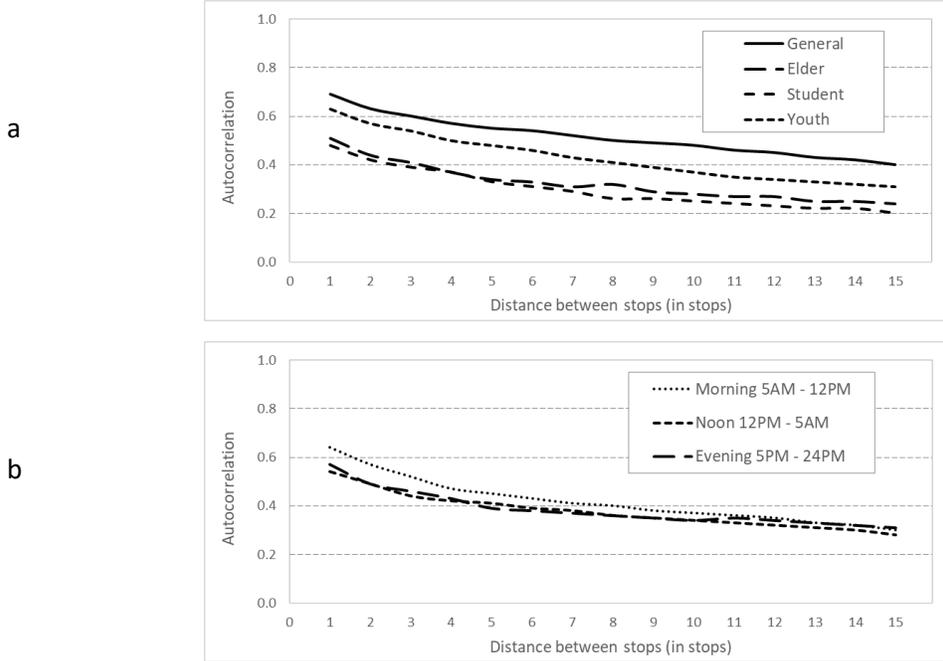

Figure 10. The correlation between the shares of occasional trips for sequential stops of the same line, by user profile (a), and by the period of the day for all profiles together (b).

### 4.5.2. Correlation between the share of occasional trips at nearby stops

To estimate the correlation between the shares $x_i$ of occasional trips at the nearby stops we applied the Moran's Index (I) of spatial autocorrelation (O'Sullivan and Unwin, 2001):

$$I = \frac{N}{W} \frac{\Sigma_i \Sigma_j (x_i - \bar{x})(x_j - \bar{x})}{\Sigma_i (x_i - \bar{x})^2} \qquad (1)$$

where $N$ is the number of stops; $x$ is the share of occasional stops, and $\bar{x}$ is the average of $x$. As a "nearby" to a stop, we consider stops at a distance less than 400 meters from a certain stop. We employ the Geoda (Anselin et al. 2010) software package for estimating Moran's I and its significance.

The value of Moran's I is positive and significant at $p < 0.01$ for all user profiles and periods of the day (Table 5), while, naturally, lower than the value of this index for the consecutive stops of the same line (Figure 10).

Table 5: Moran's I spatial autocorrelation. All R values are significant at $p < 0.01$

| User profile/period of day | Moran's I |
|---|---|
| Overall | 0.45 |
| User profile | |
| General | 0.38 |
| Youth | 0.40 |
| Elder | 0.29 |
| Student | 0.20 |
| Period of day | |
| Morning (5 AM – 12 PM) | 0.42 |
| Noon-afternoon (12 PM – 5 PM) | 0.36 |
| Evening (5 PM – 24 AM) | 0.30 |

### 4.6. Spatio-temporal pattern of occasional boardings

We investigate the spatio-temporal pattern of occasional rides for the central part of the Tel Aviv metropolitan area (Tel-Aviv Metropolitan Center, TMC) that includes Tel Aviv and four neighboring cities: Givatayim (pop. 60k), Ramat Gan (pop. 150k), Bnei Brak (pop. 190k) and Petah Tikva (pop. 230k). This area is served by ca. 600 bus lines with 1,500 stops.

The general statistics of public transport ridership within the TMC and for the rest of the county are similar (Table 6). During the day, the total share of occasional boardings within the TMC varies from 34% in the morning (5 AM – 12 PM) to 44% at noon (12 PM – 5 PM) and 54% in the evening (5 PM – 24 AM), the same as in the rest of the country.

Table 6: Ridership statistics for the Tel Aviv Metropolitan Center (Figures 11-13) and the rest of the country

| Statistic | TMC | | | The rest of the country | | |
|---|---|---|---|---|---|---|
| | Periodic Pass holders | Basic Fare Pass holders | Total | Periodic Pass holders | Basic Fare Pass holders | Total |
| Boardings | 3.6M | 5.6M | 9.2M | 8.6M | 16.9M | 25.5M |
| Unique users | 150K | 900K | 1.05 M | 279K | 1.8M | 2.1M |
| Boarding days/month | 16.0 | 6.0 | 7.4 | 15.7 | 5.7 | 7.0 |
| Boardings/month | 39.5 | 11.8 | 15.8 | 40.1 | 11.2 | 15.0 |
| Percentage of occasional boardings | 24.8% | 52.6% | 41.7% | 24.3% | 50.1% | 41.4% |

Figure 11 presents the shares and the numbers of occasional boardings at the TMC stops based on the stops' Voronoi coverage by three periods of a day. The occasional volumes are relatively steady over the day (Figure 11 d-e) compared to the shares of occasional trips (Figure 11 a-c), that grow during the day.

Let us have a close look at the areas with high shares of occasional trips > 50% in all three periods of the day (Figure 11 a-c). These include several non-residential areas of the Tel Aviv University in the northwest of the city, which may be attributed to the students' irregular ridership; the Ramat Gan – Tel Aviv Business Center, possibly indicating meetings and other business activity there and in the nearby areas with a high share of offices; the outdoor Carmel Market and nearby neighborhoods Neve Tzedek and Nahalat Binyamin - market and leisure areas with tourist attractions and numerous cafes and shops; Old Tel Aviv Port Area, another tourist area in Tel-Aviv; Sheba Medical Center in Ramat Gan.

Ramat Gan – Tel Aviv Business Center and Carmel Market area have both high rates and volumes. The Sheba Medical Center, on the other hand, exhibits only medium volumes of occasional trips despite the high share, relating to lower overall volumes of ridership in the area. Mixed-use Bnei Brak city center exhibits high volumes of occasional ridership even when the share of occasional trips is low, which can be attributed to high volumes of PT users.

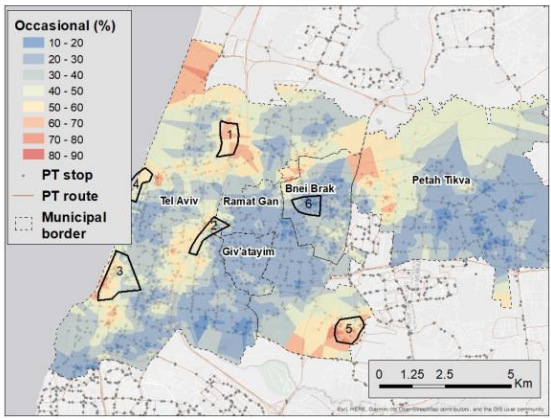 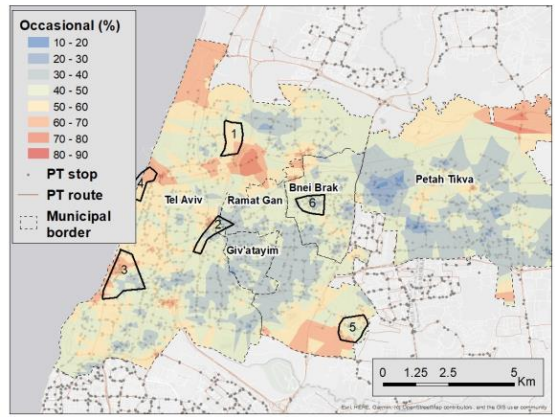 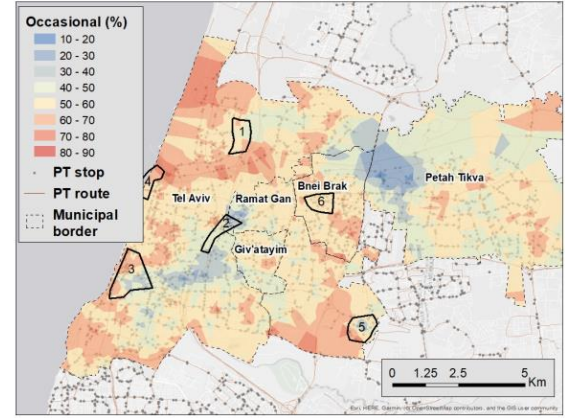

a
b
c

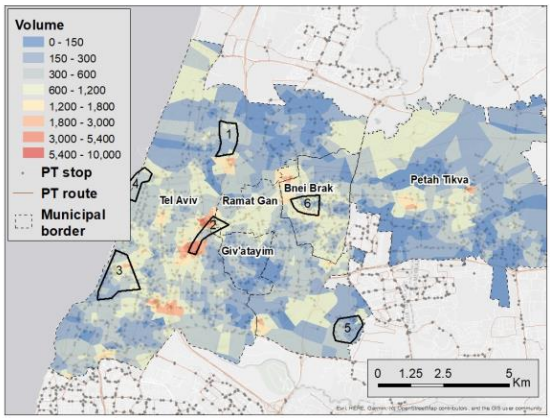 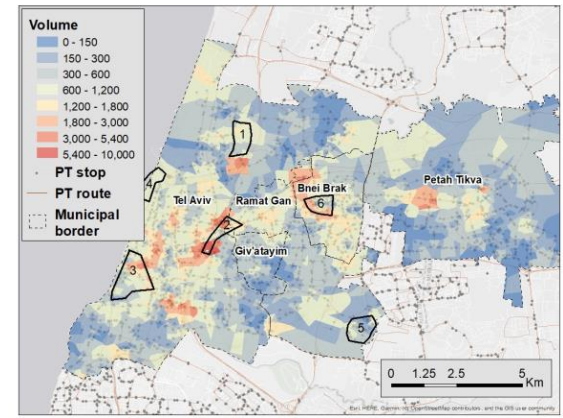 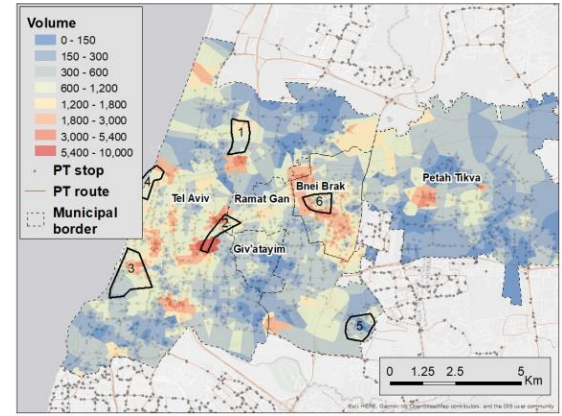

d
e
f

Figure 11: The share (top) and the number (bottom) of occasional trips at a stop (a, d) morning 5:00 – 12:00, (b, e) noon 12:00 – 17:00, (c, f) evening 17:00 – 24:00. The areas marked: (1) - Tel Aviv University; (2) Ramat Gan – Tel Aviv Business Center; (3) Carmel Market, Neve Tzedek, and Nahalat Binyamin; (4) Old Tel Aviv Port Area; (5) Sheba Medical Center in Ramat Gan; (6) Bnei Brak city center.

The formal view of the spatial and temporal correlation between the share and the number of occasional trips at a stop is presented in Figure 12. The shares of occasional boardings at two sequential hours are strongly and positively correlated with *r* ~ 0.75 (*p* < 0.01) and with the increase in time lag the correlation decreases yet remains high, indicating consistent attractiveness of stops for occasional boardings throughout the day (Figure 12a). The numbers of occasional boardings correlate even stronger, with r ~ 0.95 for the boardings at two consecutive hours, and r ~ 0.55 for a time lag of 12 hours. The spatial autocorrelation of the shares and numbers of occasional boarding is measured by the value of Moran's I between shares/numbers of boardings at a stop and the shares/numbers of boardings at neighboring stops. The value is highest in the morning and decreases with the increase in the radius of the neighborhood from r ~ 0.5 for the neighborhood of 100m radius to I ~ 0.2 for the neighborhood of 1,200m radius for the shares and from I ~ 0.25 to I ~ 0.15 for the numbers of boardings.

a
b
c
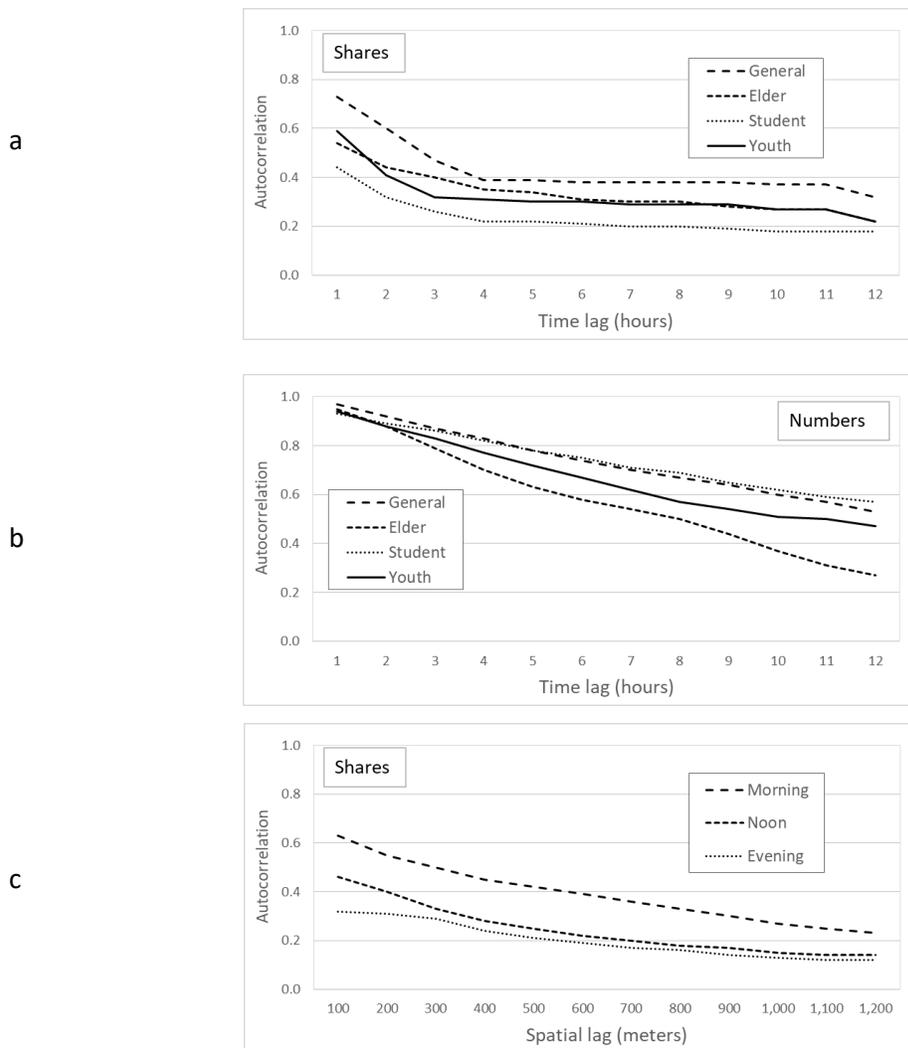

d 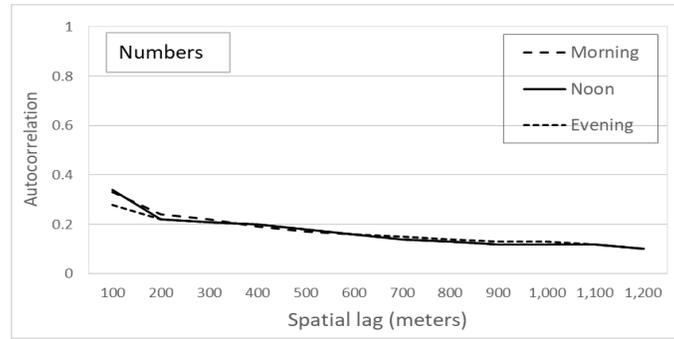

Figure 12. The temporal (a, b), estimated by travelers' profiles, and spatial (c, d) autocorrelation, estimated by the period of a day, between the shares and numbers of occasional trips at stops.

The above visual analysis and high autocorrelation between the nearby shares and volumes of occasional boardings indicate a possible relationship between the land uses and the volume of occasional boarding at the stop. To estimate this relationship, we exploited the layer of buildings of the National Geographic Data Base (BNTL) of the Survey of Israel (SOI 2018). This layer contains the foundation polygon of each building with the attributes of use (residential, commercial, industrial, public, transportation, agriculture) and building height. Based on this layer, we calculated the residential and non-residential floor area within a 400m radius of each stop (Figure 14). The correlation between the total volume of occasional trips (Figure 13) and the non-residential floor area is r = 0.3 ($p < 0.01$) and is only loosely dependent on the profiles and periods of the day (Table 7).

Table 7: The correlations between the number of occasional boardings at a stop and the size of non-residential/residential built-up area within a 400m radius around, all significant at $p < 0.01$

| User profile/period of day | Non-residential | Residential |
|---|---|---|
| Overall | 0.30 | 0.05 |
| User profile | | |
| General | 0.30 | 0.04 |
| Youth | 0.28 | 0.05 |
| Elder | 0.28 | 0.05 |
| Student | 0.21 | 0.01 |
| Period of day | | |
| Morning | 0.25 | 0.02 |
| Noon | 0.31 | 0.05 |
| Evening | 0.30 | 0.04 |

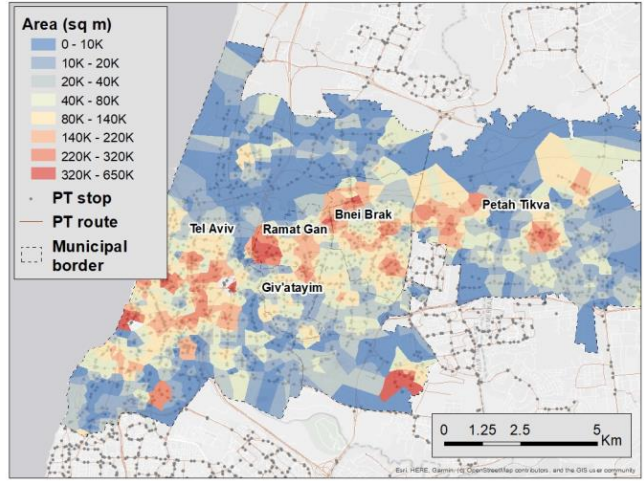

Figure 14: Voronoi diagram of the non-residential built-up area within the 400m neighborhoods around stops.

Note, that the non-residential uses are limited to specific areas that, in most cases, also include residential use. Many residential areas, however, do not include any other uses. That is why the correlation between the total volume of occasional trips and the amount of the residential floor area within the 400m neighborhood of the stop remains close to zero in all investigated cases (Table 7).

A simple linear regression provides a reasonably good estimate of the number of occasional boarding at a stop as dependent on the total number of boardings at this stop, the amount of non-residential built-up area around, and the period of the day. We construct it as:

$$y = X_1\beta_1 + X_2\beta_2 + D_1\gamma_1 + D_2\gamma_2 + e \qquad (2)$$

Where $y$ denotes the number of occasional boardings at a stop, $X_1$ represents the non-residential built-up area within the 400m neighborhood of a stop, $X_2$ is the overall number of boardings at the stop, $D_1$ and $D_2$ are dummy binary variables for noon and evening boardings, respectively, and $\beta_1$, $\beta_2$, $\gamma_1$, and $\gamma_2$ are regression coefficients.

The regression coefficients, all highly significant, are presented in Table 8, and $R^2$ = 0.85. As can be expected, the effect of the non-residential area size is positive. The residuals of (2) are slightly autocorrelated indicating that additional spatial factors can improve the prediction. The value of the Moran's I (Anselin, 2005) for the model (2) residuals, estimated for the 400m neighborhood using the *spdep R*-library (Bivand et al. 2022), is highly significant, I = 0.11, p < 0.001.

Table 8: The parameters of the linear regression (2), all coefficients significant at p < 0.01.

| | |
|---|---|
| Log-likelihood | -45,087 |
| Intercept | -74.6 |
| Total number of boardings | 0.3238 |
| Built-up area | 0.00077 |
| Morning (baseline) | ---------- |
| Noon | 203.6 |
| Evening | 274.1 |
| $R^2$ | 0.85 |
| Moran's I on residuals | 0.11 |

Figure 15 presents the maps of the observed and predicted number of occasional boardings at stops and the scatterplot of the predicted versus the observed values.

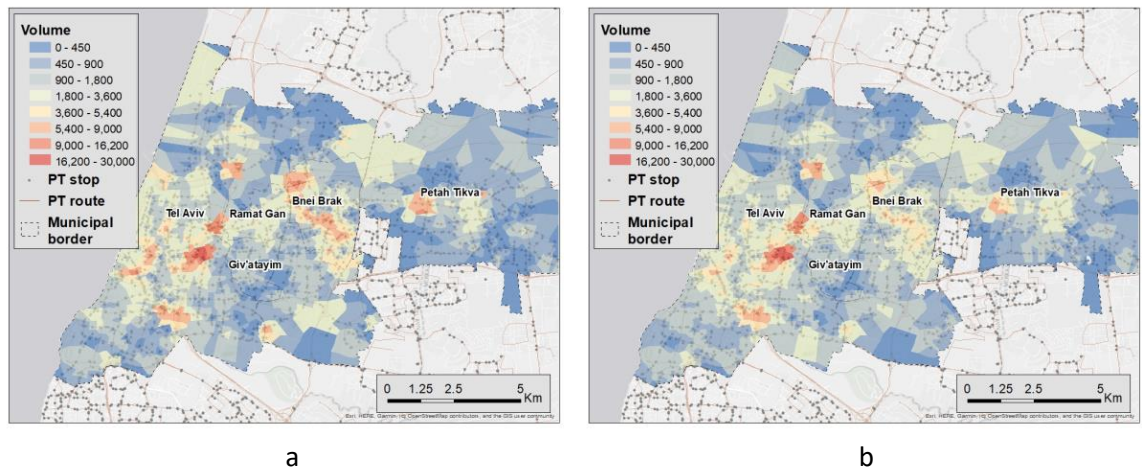

a  b

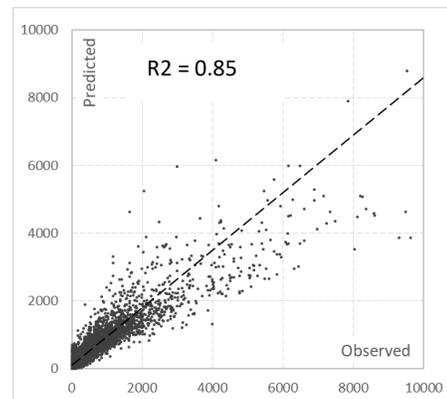

c

Figure 15: The observed (a) and predicted (b) overall numbers of occasional boardings mapped based on the Voronoi diagrams of stops, and the scatterplot of the predicted versus the observed values based on model 2 (c) (axes are truncated at 10,000).

## 5. Discussion

Demand-Responsive Transportation (DRT) attracts much of its ridership from conventional public transport (Schaller 2021). In this paper, we assume that the public transport (PT) traveler will consider DRT services for her trip when this journey is not a repeating commuting trip to work or studies, and she must plan her journey anew. We estimate the share of occasional trips in a 63 million records database of the PT boardings of the smartcard users in Israel performed during 20 working days of June 2019, which represents more than 90% of all PT trips in the working days of that month.

To estimate the volume of occasional trips, we cluster the boardings made in June 2019, based on the stop location and boarding time, and consider boardings that do not belong to any cluster as occasional. We distinguish between two categories of riders - the Basic Fare Pass (BP) users, who pay for every trip, and Periodic Pass (PP) users, who prepay for unlimited ridership for a month or more. We also distinguish between the users of several "profiles", focusing on Students, Youth, Elders, and General (all the rest).

### 5.1. Major findings

Conventional commuting is not a major part of public transport ridership. Most of the riders use public transport in a more complex way, and as many as 42% of all boardings cannot be related to any cluster of repeating trips and should be considered occasional. Specifically:

- Among 2.1 million of the monthly population of PT users, only 13% use Prepaid smartcard. These users make 34% of the trips, two boarding per working day on average. The Basic Fare Pass users made the remaining 66% of the trips and boarded, on average, four times less, 11 times in the 20 working days of July 2019.
- Occasional trips are performed by all riders. They comprise 24% of the Prepaid Pass users' trips and 51% of the Basic Fare Pass users' trips. The major reason for this gap is the difference in the monthly number of boardings between BP and PP users. The share of occasional trips decreases drastically with the increase in the total monthly number of trips for both categories, but always remains above 20%.
- The students' share of occasional trips is the highest - 37% for the PP holders and 53% for the BP holders, whereas for teenagers (the Youth profile) these shares are the lowest - 19% among the PP card holders and 49% among the BP card holders.
- Occasional boarding is a spatio-temporal phenomenon.
  - The share of occasional trips reflects the workday schedule. It is the lowest early morning (25%) and then raises twice, to 62% in the late morning, and 50% in the afternoon. The peak 75% share is characteristic of the evening.
  - The share of occasional trips essentially varies by PT line and stops. For a given line and stop, the share remains relatively steady throughout the day.
  - The number of occasional boardings at a stop is correlated with the number of occasional boardings at the stops within a 400m distance and, also, increases with the overall number of boardings within this neighborhood as well as with the non-residential built-up area.
  - A linear regression model that is based on the standard smartcard and land-use data provides a sufficiently good forecast of the number of occasional boarding at a stop.

## 5.2. Policy implications

Until very recently, the Israeli public transportation system consisted of buses and trains. Following the worldwide tendency of making PT more flexible, shared DRT services have been introduced in Tel Aviv, Haifa, and Jerusalem in 2019. To attract users during the introduction period, shared DRT modes are heavily subsidized (Globes 2019): For example, in Tel Aviv, the minibus services with flexible pick-up and drop-off locations operate at a subsidized constant price of 12.5 NIS per trip. Being twice higher than the price of a single bus trip, it is yet several times lower than the 40-80 NIS price of a typical taxi ride.

Israeli policymakers may have a hard time balancing the public demand for the shared DRT services and controlling an unwanted massive switch of the PT users. Shared DRT taxis that cancel the need for transfer between the lines and, potentially, waiting time at a stop are attractive no matter how good bus accessibility is. To repeat the USA experience, Transportation Network Companies are competing with PT (Erhardt et al. 2021), leading to a decrease in ridership by 10% between 2012-2018 (Erhardt et al. 2022) and resulting in growing Vehicle Miles Travelled (VMT) (Schaller 2021). These phenomena are strongly unwanted in the already highly congested Israeli transportation system (TomTom 2021).

One can think about decreasing the subsidies and, thus, increasing the price difference between shared DRT and buses. However, this may deter users who otherwise would gain essential accessibility surplus with the DRT. On the other hand, higher frequency and extended service hours, accompanied by targeted promotion campaigns (Cats and Ferranti 2021) demand an essential increase in subsidies. Given DRT's superiority over PT for occasional trip-making, the only robust policy direction seems to be the reduction of private car use which remains the main contributor to traffic congestion (CBS 2020). DRT may not be able to compete with private cars if drivers are unwilling to give up their privacy, as demonstrated in the US (Lavieri and Bhat 2019) and Australia (Krueger et al. 2016). Yet in the Israeli context, most of the 2.1M Basic Fare Pass cardholders have access to other modes of transport including private cars and the policymaker may attempt to persuade them to shift some of their trips with the private car to DRT. Possible policy tools for that can be locally adjusted demand-responsive parking prices (Fulman and Benenson 2019) or congestion pricing (Ben-Dor et al. 2022). Many features make DRT modes appealing for occasional trips, and in cities with prominent private DRT ridership, decision-makers should prioritize shared DRT modes.

With shared DRT consuming the demand for occasional trips in city centers, operators of conventional PT will be able to focus on commuting trips. The lines that attempt to serve many different activities could be simplified, while the lines that mostly serve the occasional ridership can be canceled and resources diverted to increase the frequency of PT and improve the level of service for commuters. In Barcelona, a major simplification of the bus network that took place between 2012-2018 allowed a reduction of the average headway by half, from 12.3 to 6.2 minutes, and led to an increase in demand (Badia et al. 2017). In Israel, private car owners may switch to PT for commuting if the conventional system is improved. To conclude, left to the market forces, modern DRT has a high potential to dry out conventional PT services and further worsen the state of the transportation system. Our study points to the PT user groups that should be the focus of the policy regulations that would prevent this.

## Acknowledgements

The research was funded by the research grant "Estimating the demand for public transport based on the smartcard and mobile phone data" #0606915691 of the Israeli Ministry of Transport and Road Safety (MOT).

The authors are grateful to Mr. Zeev Shadmi, MOT, Ms. Sarit Levy, MOT, and Mr. Doron Narkis, MOT advisor, for their great support in smartcard data querying and management.

# Appendix: The robustness of the extended DBSCAN

Our analysis in the main text is based on the spatio-temporal DBSCAN with the following parameters minPts = 2, $\epsilon_s$ = 400, $\epsilon_t$ = 60. Here we investigate the robustness of results of the clustering procedure to these parameters varying these parameters within the following limits: minPts = 2 and 3, $\epsilon_t$ = 60 and 120 minutes, and $\epsilon_s$ = 250 and 400 m (Table A1). The grades of the $\epsilon_s$ are chosen based on the estimates of the walking distance to the nearest PT stop (Canepa 2007).

Table A1: DBSCAN parameters for sensitivity analysis, and the share of occasional boardings

| Set | minPts | $\epsilon_s$ | $\epsilon_t$ | Percentage occasional |
|---|---|---|---|---|
| A (original) | 2 | 400 | 60 | 42% |
| B | 2 | 400 | 120 | 34% |
| C | 2 | 250 | 60 | 46% |
| D | 3 | 400 | 60 | 54% |
| E | 3 | 400 | 120 | 46% |
| F | 3 | 250 | 60 | 58% |

The trends of ridership regularity (Figures A1-A3) repeat the results obtained for the base set of parameters - minPts = 2, $\epsilon_s$ = 400, $\epsilon_t$ = 60 (see Figures 5, 6, and 8 in the major text). With the increase in the user's number of boarding, the share of occasional trips decreases yet remains above 20% for all parameters' sets (Figure A1); the number of clusters increases (Figure A2). The share of occasional trips always remains lowest in the morning and afternoon (Figure A3).

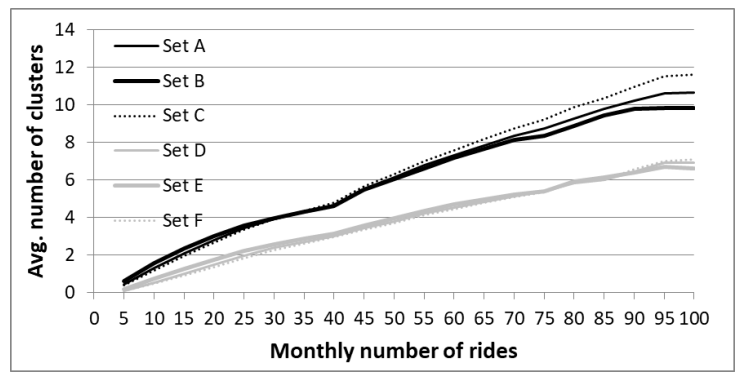

Figure A1: Share of occasional rides as dependent on the number of boardings made in working days of June, for the different DBSCAN parameter sets (see Table 8).

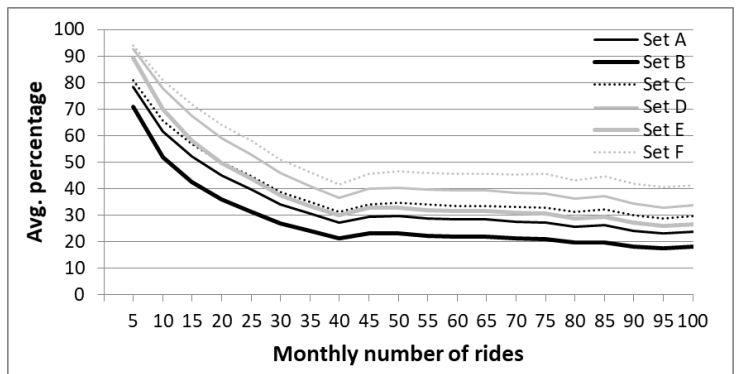

Figure A2: Average number of clusters as dependent on the number of boardings made in working days of June, for the different DBSCAN parameter sets (see Table 8).

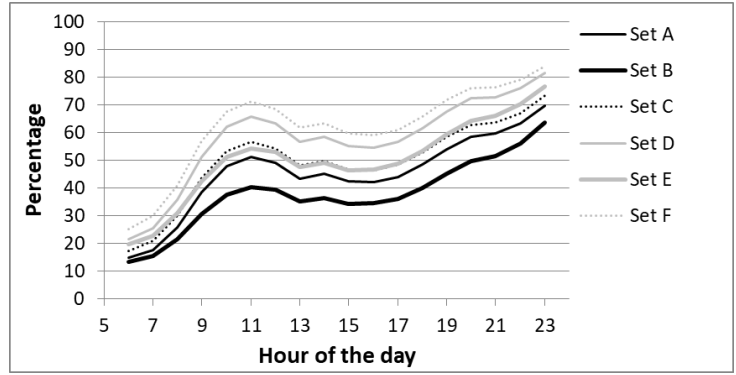

Figure A3: Share of occasional rides by the hour of the day, for the different DBSCAN parameter sets (see Table 8).

The decrease of $\epsilon_s$ from 400m to 250m consistently leads to an increase in the percentage of occasional trips (Figure A1), while increasing $\epsilon_t$ from 60 to 120 results in fewer boardings classified as occasional. The minPts parameter has, unsurprisingly, the most significant impact on results. The step form minPts = 2 to 3 'converts' some of the clustered trips into non-clustered and minPts = 3 raises the average occasional ridership from the baseline of 42% to 54%. The average number of clusters (Figure A2) remains the same for different distance and time parameters.